\documentclass[conference]{IEEEtran}
\IEEEoverridecommandlockouts
\usepackage{cite}
\usepackage{amsmath,amssymb,amsfonts}
\usepackage{algorithmic}
\usepackage{graphicx}
\usepackage{textcomp}
\usepackage{xcolor}
\usepackage{hyperref}
\usepackage{tabularx}
\usepackage{booktabs}
\usepackage{orcidlink}

\makeatletter
\newcommand{\printfnsymbol}[1]{%
  \textsuperscript{\@fnsymbol{#1}}%
}
\makeatother

\def\BibTeX{{\rm B\kern-.05em{\sc i\kern-.025em b}\kern-.08em
    T\kern-.1667em\lower.7ex\hbox{E}\kern-.125emX}}
\begin{document}
\title{\textit{Divide and Conquer:} A Large-Scale Dataset and Model for Left–Right Breast MRI Segmentation}
\author{
\IEEEauthorblockN{Maximilian Rokuss\textsuperscript{1,3,5}\orcidlink{0009-0004-4560-0760}\printfnsymbol{1}\thanks{\printfnsymbol{1}\small Co-first authors may list themselves as lead author on their CV.}, 
Benjamin Hamm\textsuperscript{1,4}\orcidlink{0009-0003-4818-8700}\printfnsymbol{1}, 
Yannick Kirchhoff\textsuperscript{1,3,5}\orcidlink{0000-0001-8124-8435}\printfnsymbol{1}, 
Klaus Maier-Hein\textsuperscript{1,2,3,4,5}\orcidlink{0000-0002-6626-2463}}
\vspace{0.2cm}
\IEEEauthorblockA{\textsuperscript{1}German Cancer Research Center (DKFZ) Heidelberg, Division of Medical Image Computing, Germany}
\IEEEauthorblockA{\textsuperscript{2}Pattern Analysis and Learning Group, Department of Radiation Oncology, Heidelberg University Hospital}
\IEEEauthorblockA{\textsuperscript{3}Faculty of Mathematics and Computer Science, Heidelberg University}
\IEEEauthorblockA{\textsuperscript{4}Medical Faculty, Heidelberg University, Germany}
\IEEEauthorblockA{\textsuperscript{5}HIDSS4Health, Karlsruhe/Heidelberg, Germany}
\IEEEauthorblockA{\tt\small\{maximilian.rokuss, benjamin.hamm, yannick.kirchhoff\}@dkfz-heidelberg.de}
\vspace{0.2cm}
\IEEEauthorblockA{\tt\small\url{www.github.com/MIC-DKFZ/BreastDivider}}
}

\maketitle
\section*{Introduction}

\noindent Breast cancer continues to be a leading cause of morbidity and mortality among women worldwide. Magnetic Resonance Imaging (MRI) is a cornerstone in the detection, diagnosis, and treatment planning of breast cancer, offering high-resolution, non-invasive visualization of breast tissue. Recent advances in medical imaging and machine learning have significantly improved automated analysis of breast MRI data, enabling more precise and scalable diagnostic tools~\cite{classical_seg,breast_classify}. However, a critical gap remains: existing algorithms often fail to explicitly segment and distinguish the left and right breasts as separate regions of interest (ROIs). Models designed to operate on a per-breast basis may struggle in cases where both breasts are imaged or where one is absent, for instance, following mastectomy.\\

\noindent Precisely localized ROIs—such as clear separation of the left and right breasts—can support a wide range of tasks beyond segmentation, including classification~\cite{breast_classify}, prompt-based lesion localization~\cite{lesionlocator}, and structured report generation~\cite{chen2025large}. In dynamic contrast-enhanced MRI, where multiple post-contrast images are acquired following a baseline scan, a reliable ROI from the first timepoint can serve as a spatial prior across the entire temporal sequence. This enables temporally informed segmentation~\cite{longiseg} while reducing the need to segment each full volume individually, improving both efficiency and consistency. At the same time, large annotated datasets are being actively leveraged for pretraining deep learning models in medical imaging~\cite{suprem}. By supporting both region-based modeling and scalable pretraining, detailed ROI annotations on a large-scale dataset can serve as a foundational asset across diverse downstream applications.\\

\noindent While resources such as the Duke-Breast-Cancer-MRI dataset~\cite{duke_breast_mri} and MAMA-MIA~\cite{mama_mia_dataset} have laid the groundwork by providing limited whole breast segmentation masks, and classical image processing methods can offer rough ROI estimations~\cite{classical_seg}, the field still lacks a large-scale, annotated dataset that provides detailed, side-specific breast segmentations at a dataset size which is also suitable for model pretraining.\\

\begin{figure}[t]
  \centering
  \includegraphics[width=\linewidth]{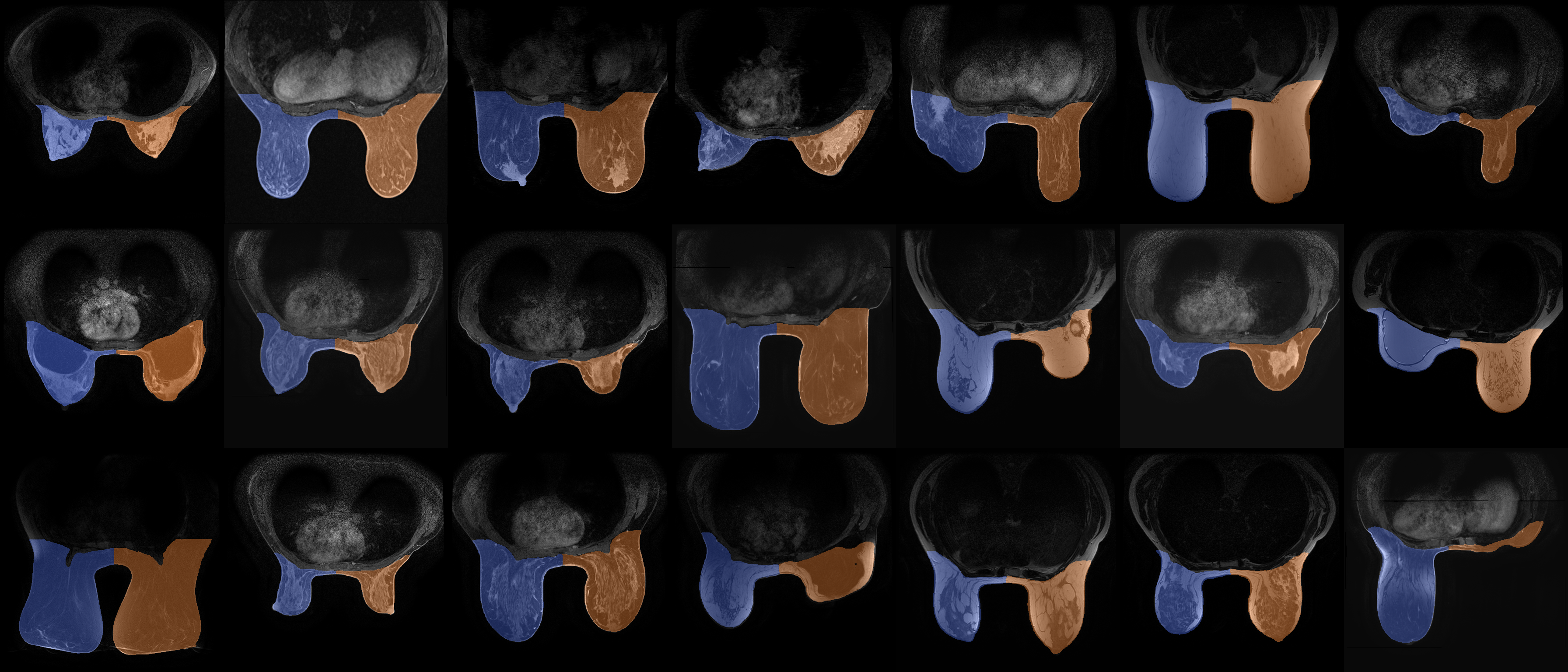}
  \caption{Example axial breast MRI slices from the compiled dataset with overlaid segmentation masks. The left (blue) and right (orange) breast regions are shown, demonstrating consistent anatomical separation and high-quality annotations across diverse imaging sources. Especially the last column highlights challenging edge cases including \textit{breast asymmetry}, the presence of \textit{breast implants}, and \textit{unilateral breast absence following mastectomy}.}
  \vspace{-0.3cm}
  \label{fig:qualitative}
\end{figure}

\noindent To address this need, we introduce the first publicly available, large-scale breast MRI dataset with explicit left and right breast segmentation labels. Qualitative examples are illustrated in Fig.~\ref{fig:qualitative}. The dataset comprises over 13,000 annotated 3D MRI volumes, making it \textit{one of the largest segmentation datasets in medical imaging} to date. In addition, we release a robust deep learning model based on nnU-Net~\cite{isensee2021nnu}, trained specifically for left-right breast segmentation. Our goal is to establish a foundational resource that enables fine-grained breast MRI analysis, supports downstream tasks such as tumor detection and classification, and facilitates a divide-and-conquer strategy by providing high-quality ROIs.\\

\noindent This contribution is notable not only for its scale and quality but also for its diversity and standardization. The dataset aggregates MRI scans from a broad range of publicly available sources, all of which have been harmonized to a consistent RAS (Right-Anterior-Superior) orientation and converted into the NIfTI format. This standardization ensures spatial consistency across the dataset and significantly reduces the preprocessing burden for researchers. Moreover, the dataset’s size makes it a strong candidate for use in \textit{pretraining large-scale models}. By providing both imaging data and comprehensive segmentation labels, this resource serves as a high-quality benchmark and a practical tool for advancing research in women’s health and breast imaging. The dataset and trained model are publicly available at: \url{www.github.com/MIC-DKFZ/BreastDivider}

\section*{Materials and Methods}

\subsection*{Dataset Compilation}


\noindent We curated a diverse and comprehensive breast MRI dataset by aggregating scans from multiple publicly available sources: the Duke-Breast-Cancer-MRI dataset \cite{duke_breast_mri}, MAMA-MIA \cite{mama_mia_dataset}, Advanced-MRI-Breast-Lesions \cite{daniels2024advanced}, and EA1141 \cite{comstock2023ea1141}, totaling 13,752 3D scans. The dataset includes a wide range of common MRI modalities such as T1-weighted, T2-weighted, T1 with contrast (T1+C), and FLAIR, reflecting the heterogeneity of clinical imaging protocols. The Advanced-MRI-Breast-Lesions dataset (single-institutional) incorporated T1 DCE with five fat-saturated phases, delayed T1 (fat saturated), T2, and T1 (non-fat saturated) sequences. The Duke-Breast-Cancer-MRI dataset (single-center) includes pre-operative dynamic contrast-enhanced (DCE) MRI acquired at 1.5T or 3T, comprising non-fat saturated T1-weighted sequences, fat-saturated gradient echo T1-weighted pre-contrast images, and typically three to four post-contrast sequences. The EA1141 dataset, collected across 48 different clinical sites, provides non-contrast (pre-contrast) T1-weighted, first post-contrast T1-weighted, T2-weighted, and diffusion-weighted imaging (DWI), with some cases including a second post-contrast T1 sequence. To ensure consistency and maintain a minimum quality threshold, only MRI volumes with at least 32 slices per axis and a resolution of 3\verb#×#3\verb#×#3 mm or finer were included. From the Duke dataset, we selected 100 cases with existing whole-breast segmentation labels, which served as the foundation for developing our initial model.

\subsection*{Left-Right Breast Label Generation}


\noindent To generate left and right breast segmentation labels, we employed a center-of-mass (COM) splitting technique applied to the initial 100-case whole-breast segmentation dataset. For each 3D MRI scan, we first computed the centroid of the segmented breast tissue by taking the voxel-wise average of all coordinates within the binary mask, resulting in a single center-of-mass point in image space. We then identified the sagittal splitting plane that intersects with the centroid. The original breast mask was subsequently divided into two parts by assigning all voxels to the left or right breast based on their position relative to this plane. Prior to model training, all generated left–right labels were visually inspected to confirm anatomical plausibility and correct assignment.\\

\subsection*{Model Training and Active Learning}

\noindent We utilized the nnU-Net framework \cite{isensee2021nnu} in a low-resolution configuration to accommodate both breasts in the patch size of the network when given high-resolution MRI scans. The initial model was trained on a 100-case dataset with the above described generated left–right labels. To improve performance and generalizability, we employed an active learning strategy that leveraged predictions from non–contrast-enhanced T1-weighted images as pseudo-ground truth for co-registered sequences, building on previously demonstrated reliability of cross-sequence annotation transfer~\cite{cross-seq}. The trained model was applied to the remaining unlabelled cases, and the predictions were refined through an iterative loop of inspection and retraining. High-confidence predictions were incorporated into the training set for the next training round, ultimately producing a final dataset of 13,752 cases with high-quality left–right breast segmentation labels.

\subsection*{Manual Inspection and Quality Control}

\noindent All predicted segmentations underwent visual inspection by a team of three experts. The dataset was split evenly among them to ensure that 100\% of the cases were reviewed. Each case was either accepted as-is, discarded if the input images were broken or of insufficient quality, or flagged for refinement if the segmentation quality was inadequate. Refinements were performed either manually using nnInteractive~\cite{nninteractive} or by including the case in the next iteration of the active learning loop. This comprehensive review process ensured that all included segmentations met a consistent quality standard prior to inclusion in the final dataset.

\section*{Results and Discussion}


\begin{table}[t]
\centering
\begin{tabular}{cccc}
\toprule
Metrics & DSC [\%] & NSD@1mm [\%] & HD95 [mm] \\
\midrule
Fold 0 & $99.11\pm0.45$ & $99.12\pm0.94$ & $0.0210\pm0.1657$ \\
Fold 1 & $99.05\pm0.48$ & $99.07\pm0.97$ & $0.0270\pm0.3243$ \\
Fold 2 & $99.08\pm0.46$ & $99.10\pm0.91$ & $0.0167\pm0.1120$ \\
Fold 3 & $99.05\pm0.60$ & $99.08\pm0.97$ & $0.0248\pm0.1381$ \\
Fold 4 & $99.1\pm0.40$ & $99.15\pm0.80$ & $0.0086\pm0.0804$ \\
\midrule
Mean & $99.08\pm0.48$ & $99.10\pm0.92$ & $0.0196\pm0.1849$ \\
\bottomrule
\end{tabular}
\vspace{0.2cm}
\caption{}
\vspace{-0.3cm}
\begin{quote}
\footnotesize Performance metrics for each fold in a 5-fold cross-validation. Results are reported as mean $\pm$ standard deviation for the Dice Similarity Coefficient (DSC), Normalized Surface Dice at 1mm (NSD@1mm), and the 95th percentile Hausdorff Distance (HD95). The model demonstrates high consistency and robustness across folds, with high average scores reflecting the model’s strong performance, aided by the large dataset and consistent, relatively large target structures (left and right breasts).
\end{quote}
\vspace{-0.5cm}
\label{tab:folds}
\end{table}

\noindent To assess the performance and generalizability of our nnU-Net model for left and right breast segmentation in MRI scans, we conducted a 5-fold cross-validation. The dataset was partitioned into five subsets, ensuring that each fold served as a validation set once while the remaining four were used for training. The segmentation accuracy was evaluated using the Dice Similarity Coefficient (DSC), a standard metric for measuring the overlap between the predicted segmentation and the ground truth. The results across the five folds are summarized in Table \ref{tab:folds}. These results indicate that the model consistently achieved high segmentation accuracy across all folds, with minimal variance, demonstrating its robustness and reliability. The high DSC values underscore the model's capability to accurately delineate breast tissues, which is crucial for downstream tasks such as tumor detection, treatment planning, and monitoring.\\ 

\noindent The availability of this large-scale, publicly accessible dataset with explicit left-right breast segmentation labels addresses a significant gap in breast MRI research. It enables the development and benchmarking of algorithms that require individual breast analysis, such as those for detecting unilateral abnormalities or assessing asymmetry. Moreover, the dataset supports research into cases involving mastectomy, where traditional per-breast models may fail.

\noindent By providing both the dataset and the trained nnU-Net model to the research community, we aim to catalyze advancements in breast cancer imaging analysis. This resource facilitates the development of more sophisticated and clinically applicable MIC and CAI tools, ultimately contributing to improved diagnostic accuracy and patient outcomes in women's health.

\section*{Conclusion}

\noindent We have introduced the first publicly available breast MRI dataset with explicit left and right breast segmentation labels, encompassing more than 13k annotated cases. Alongside this dataset, we provide a robust deep-learning model trained for left-right breast segmentation. This work addresses a critical gap in breast MRI analysis and offers a valuable resource for the development of advanced tools in women's health.

\vspace{0.5cm}
\section*{Acknowledgement}

\noindent The present contribution is supported by the Helmholtz Association under the joint research school "HIDSS4Health – Helmholtz Information and Data Science School for Health".

\newpage
\bibliographystyle{IEEEtran}
\bibliography{bibliography}
\end{document}